\documentstyle{natokap}

\begin{opening}
\title{ON THE UNITARIZATION OF HIGHEST WEIGHT REPRESENTATIONS 
FOR AFFINE KAC-MOODY ALGEBRAS}
\author{JUAN GARCIA ESCUDERO and MIGUEL LORENTE}
\institute{Departamento de F{\'\i}sica\char44 Universidad de Oviedo\\
 33007\char44 Oviedo\char44 Spain}
\end{opening}

\begin{document} 

\begin{abstract}
In a recent paper ([1],[2]) we have classified explicitely all the unitary highest weight
representations of non compact real forms of semisimple Lie Algebras on Hermitian symmetric
space. These results are necessary in order to construct all the unitary highest weight
representations of affine Kac-Moody Algebras following some theorems proved by Jakobsen and Kac
([3],[4]). \end{abstract}

\section{The Affine Kac-Moody Algebra}
Let $\dot g$ be a finite dimensional semisimple complex Lie algebra with Chevalley
basis $\left\{H_\alpha, E_\alpha , F_\alpha\right\}$ with $\alpha$ belonging to the
set of simple roots. The elements of the so-called Cartan matrix $A$ are defined by
$$
{A}_{jk} ={ \alpha }_{k}\left({{H}_{\alpha j}}\right) ={2\left({{
\alpha }_{k} ,{ \alpha }_{j}}\right) \over \left({\alpha }_{j} ,{ \alpha
}_{j}\right)} ,\qquad  j , k =1,2,\ldots , \ell $$
where, $\left(\alpha_j, \alpha_k\right) = B\left(H_{\alpha_j}, H_{\alpha_k}\right),
\; B(\:,\:)$ being the Killing form of $\dot g$ and $H_\alpha$ an element of the
Cartan subalgebra $h$ which is assigned uniquely to each root $\alpha \in \Delta$ by
the requirement that $B\left(H_\alpha, H\right)=\alpha (H)$ for all $H \in h$.

The elements in the Chevalley basis satisfy
$$
\begin{array}{ll}
\left[H_{\alpha_j}, E_{\alpha_k}\right]=A_{jk}E_{\alpha_k}\; , 
&\left[H_{\alpha_j}, F_{\alpha_k}\right]= - A_{jk}F_{\alpha_k}
\\ \noalign{\smallskip}
\left[E_{\alpha_j}, F_{\alpha_k}\right]= \delta_{jk}H_{\alpha_k}\; , 
&\mbox{for all}\quad \alpha_j,\alpha_k \in \Delta
\end{array} 
$$

Every finite dimensional semisimple complex Lie algebra can be constructed from its
Cartan matrix $A$ which satisfies the following propierties.
\begin{enumerate}
\item[a)]
 ${A}_{ii}=2 \quad \forall i, \quad i=1,\ldots ,\ell$
\item[b)]
${A}_{ij}=0,-1,-2, \quad {\rm or} \; -3 \; {\rm if} \; i\ne j, \quad i,j
=1,\ldots ,\ell $
\item[c)]
${A}_{ij}=0 \; \mbox{ if and only if} \; {A}_{ji}=0$
\item[d)]
det $A$ and all proper principal minors of $A$ are positive.  
\end{enumerate}

The starting point on the construction of infinite dimensional Kac-Moody algebras is
the definition of a generalized Cartan-matrix with elements $A_{ij}$ ($i,j\in I, \;
I=0,1, \ldots ,\ell $) satisfying
\begin{enumerate}
\item[a)]
 $A_{ii}=2 \qquad \forall i \in I$
\item[b)]
$A_{ij}$ is either zero or a negative integer for $i \neq j$.
\item[c)]
$A_{ij}=0$ if and ony if $A_{ji}=0$.
\end{enumerate}

A Lie algebra whose Cartan matrix is a generalized Cartan matrix is called a
Kac-Moody algebra ([5],[6]). A Kac-Moody algebra is affine if its generalized Cartan
matrix is such that det $A = 0$ and all the proper principal minors of $A$ are
positive. In the following we restrict ourselves to the affine case.

For a Kac-Moody algebra the Cartan subalgebra $h$ is divided into two parts
$h=h'\oplus h''$.

The basis elements of $h'$ are $H_{a_j}\left(j \in I\right)$ and $h''$ is the
one-dimensional complementary subspace of $h'$ in $h$ generated by one element 
that we call $d$.

The center $C$ in the affine case is one-dimensional. Every element of $C$ is a
multiple of $h_\delta$ where $\delta$ is defined by the following conditions
$$
\begin{array}{lll}
\delta(h) &= 0 & \quad {\rm for} \: H \in h'\\
\delta(d) &= 1 & \quad {\rm for} \: d \in h'' 
\end{array} 
$$

The Cartan subalgebra $h$ has dimension $\ell + 2$. In order to obtain a basis for $h^*$ 
we need $\ell + 2$ linear functionals. We can take as basis $\alpha_k \; (k = 0,1,
\ldots ,\ell)$, and the linear functional $\Lambda_0$ defined as:
$$
\begin{array}{ll}
\left({{\Lambda }_{0},{\alpha }_{k}}\right) &= 
\left\{\begin{array}{ll}
{1 \over 2}\left({{\alpha }_{0},{\alpha }_{0}}\right) & {\rm if} \; k=0 \\
0 & {\rm if} \; k=1,2, \ldots , \ell \end{array}\right.\\
\left({{\Lambda }_{0},\delta }\right) &= {1 \over 2}\left({{\alpha
}_{0},{\alpha }_{0}}\right)
\end{array} 
$$

There exists two types of affine complex Kac-Moody algebras: Untwisted and Twisted.
The untwisted ones $g^{(1)}$ may be constructed starting from any simple complex 
Lie algebra. The twisted affine Kac-Moody algebras $g^{(q)}$ $(q = 2,3)$ can all be
constructed as subalgebras of certain of these untwisted algebras.

\medskip
{ \noindent \bf i) Untwisted affine Kac-Moody algebras}

Let {\it \.g} be a simple complex Lie algebra of rank $\ell$. A realization of the complex
untwisted affine Kac-Moody algebra $g^{(1)}$ is given by
$${g}^{\left({1}\right)}=\rlap/C c\oplus \rlap/C d\oplus
\sum\nolimits\limits_{j\in Z} {z}^{j}\otimes \dot{g}$$

\vskip -8pt
{\noindent with the following conmutation relations}
$$
\begin{array}{l}
\left[{{z}^{j}\otimes a,{z}^{k}\otimes b
}\right]={z}^{j+k}\otimes \left[{a,b}\right]+
j{\delta }_{j,-k}B\left({a,b}\right)c \quad \forall
a,b \in {\dot{g}} 
\\ \noalign{\smallskip}
\left[z^j \otimes a,c \right] = 0 
\\ \noalign{\smallskip}
\left[ d,z^j \otimes a  \right]= jz^j \otimes a 
\\ \noalign{\smallskip}
\left[ d,c \right] = 0
\end{array}
$$

{ \noindent \bf ii) Twisted Affine Kac-Moody algebras}

Let {\it \.g} be a simple complex Lie algebra and $\tau$ a rotation of the set of
roots of {\it \.g}. If the rotation $\tau$ is not an element of the Weyl group of {\it
\.g} then there exist an associated outer automorphism $\Psi_\tau$ such that
$\Psi_\tau \left(h_\alpha\right)=h_{\tau(\alpha)}$. We have $\tau^q=-1$ and also
$\left(\Psi_\tau\right)^q=1$ with $q =2,3$. The eigenvalues of $\Psi_\tau$ are 
$e^{2\pi ip / q}, \quad p=0,1,\ldots ,q-1$. Let ${\dot{g}}_{p}^{\left({q}\right)}$ be
the eigenspace corresponding to the eigenvalue $e^{2\pi ip/ q}$. The 
Twisted Affine
Kac-Moody algebra is then:
$${g}^{\left({q}\right)}=\left({\rlap/C c}\right)\oplus \left({\rlap/C
d}\right)\oplus \sum\nolimits\limits_{p=0}^{q-1}
\sum\nolimits\limits_{\stackrel{j}{j \mbox{\scriptsize mod}\; q\;=\;p}}
\left({{z}^{j}\otimes {\dot{g}}_{p}^{(q)}}\right)$$

\section{Highest Weight Representations. The Contravariant Hermitian Form}

A subset $\Delta_+$ of $\Delta$ is called a set of positive roots if the following
propierties are satisfied

\begin{enumerate}
\item[i)] If $\alpha, \beta \in \Delta_+$ and $\alpha+\beta \in \Delta$ then
$\alpha+\beta \in \Delta_+$
\item[ii)] If $\alpha \in \Delta$ then either $\alpha$ or -$\alpha$ belongs to
$\Delta_+$ 
\item[iii)] If $\alpha \in \Delta_+$ then $-\alpha \notin \Delta_+ $
\end{enumerate}

For each set of positive roots we may construct a Borel subalgebra $b=$\hskip -7truemm
\vtop{\hsize=2,1truecm {\box1 $\hskip 6truemm\oplus$} \par \vskip -3truemm \box2 ${\scriptstyle \alpha
\in \Delta_+\cup\left\{0\right\}}$} $g_\alpha$ . A subalgebra $p \subset g$ such that $b
\subset p$ is called a parabolic subalgebra.

Let $U(g)$ denote the universal enveloping algebra of $g$ and let $\omega$ be an antilinear anti-involution of $g$ $\left({{\rm i.e.} \; \omega\left[{x,y}\right]=\left[{\omega y,\omega
x}\right] \quad {\rm and} \; \omega  \left({\lambda x}\right)=\overline{\lambda
}\omega  \left({x}\right)}\right)$ such that
$$g = p + \omega p$$

An antilinear anti-involution $\omega$ of $g$ is called consistent if $\forall 
\alpha \in \Delta , \; \omega {g}_{\alpha }={g}_{
-\alpha }$. Now let $e_i, \; f_i, \; h_i \quad (i=0, \ldots , l)$ be defined as ([5])
$$
\begin{array}{lll}
{e}_{0}=z\otimes {F}_{{\gamma }_{r}} & \quad {f}_{0}={
z}^{-1}\otimes {E}_{{\gamma }_{r}} & \quad {h}_{0}={2 \over
\left({{\gamma }_{r},{\gamma }_{r}}\right)}
c-1\otimes {H}_{{\gamma }_{r}} \\ 
{e}_{i}=1\otimes {E}_{i} & \quad {f}_{i}=1\otimes {F}_{i} & \quad {
h}_{i}=1\otimes {H}_{i} \quad i=1,\ldots ,\ell
\end{array}
$$
where $\gamma_r$ is the highest positive root.

When $\omega e_i = f_i$ and $\omega h_i =h_i \quad (i = 0,\ldots ,l)$ then $\omega$ is called the compact antilinear anti-involution and is denoted by $w_c$.

Let now $\Lambda :p\rightarrow \rlap/c $ be a 1-dimensional
representation of $p$. A representation $\Pi :g\rightarrow g\ell 
\left({V}\right)$ is called a highest weight representation with
highest weight $\Lambda$ if there exists a vector ${\vartheta }_{\Lambda }\in 
V$ satisfying
\begin{enumerate}
\item[a)] $\Pi \left({u\left({g}\right)}\right){\vartheta
}_{\Lambda }=V$
\item[b)] $\Pi \left({x}\right){\vartheta }_{\Lambda }=\Lambda 
\left({x}\right){\vartheta }_{\Lambda }\forall x\in p$
\end{enumerate}

An Hermitian form $H$ on $V$ such that
$$
\begin{array}{l}
H\left({{\vartheta }_{\Lambda },{\vartheta }_{\Lambda
}}\right) =1 \\ \noalign{\smallskip}
H\left({\Pi \left({x}\right)u,v}\right)=H\left({
u,\Pi \left({\omega x}\right)v}\right) \quad \forall x\in 
g; \quad u,v\in V
\end{array}
$$
is called contravariant. When $H$ is positive semi-definite, $\Pi$ is said to be
unitarizable.

In the following we construct the Hermitian form $H$ ([3]). We choose a subspace $n
\subset g$ such that $g=p \oplus n$. Then we have $U(g)=nU(g) \oplus U(p)$. Let
$\beta$ be the proyection on the second sumand. Let $\Lambda$ be a 1-dimensional
representation of a parabolic subalgebra $p$ (in particular a Borel subalgebra $b$ as
in the integrable representations case) satisfying $\Lambda \left({\beta 
\left({u}\right)}\right)=\overline{\Lambda \left({\beta 
\left({\omega u}\right)}\right)} \quad \forall u\in U
\left({g}\right)$.

Let ${p}^{\Lambda }=\left\{{x\in p/\Lambda \left({x
}\right)=0}\right\}$. The space
$$
{M}_{p,\omega }\left({\Lambda }\right)=U\left({
g}\right)/U\left({g}\right){p}^{\Lambda }
$$
defines a representation of $g$ on ${M}_{p,\omega }\left({\Lambda
}\right)$ via left multiplication that is called a (generalized) Verma module and that
is a highest weight representation. In addition it can be shown that there exists a
unique contravariant hermitian form defined by
$$
H\left({u,v}\right)=\Lambda \left({\beta \left({\omega
\left({v}\right)}\right)u}\right) \; {\rm for} \: u,v\in U
\left({g}\right)
$$
which is independent of the choice of $p$.

Let $I(\Lambda)$ denote the Kernel of $H$ on ${M}_{p,\omega }\left({\Lambda
}\right)$. Then $H$ is nondegenerate on the highest weight module
$$
{L}_{p,\omega }\left({\Lambda }\right)=M\left({
\Lambda }\right)/I\left({\Lambda }\right)
$$
In the following we will give for each of the unitarizable representations
(integrable, elementary and exceptional) the choice of $p$ and $\omega$ for which the hermitian form is nondegenerate and positive definite.

\section{Integrable representations}

Let ${\Pi }^{st}=\left\{{{\alpha }_{0},\ldots {\alpha }_{\ell
}}\right\}$ be the standard set of simple roots. The standard set of positive roots
([4]) is ${\Delta }_{+}^{st}=\left\{{\sum\nolimits\limits_{} {
k}_{i}{\alpha }_{i}/{k}_{i}=0,1,2,\ldots \: ; \: {\alpha }_{i
}\in {\Pi }^{st}}\right\}$ and the corresponding Borel subalgebra is denoted by
$b^{st}$:
\begin{eqnarray*}
\lefteqn{{b}^{st}=\rlap/C c\oplus \left({1\otimes \dot{
b}}\right)\oplus \left({z\otimes \dot{g}}\right)\oplus
\left({{z}^{2}\otimes \dot{g}}\right)\oplus \ldots}\\
\noalign{}
&\quad & = \; \mbox{span}\;\left\{{{z}^{k}\otimes {h}_{i}/k\ge 0,i=0,\ldots 
\ell }\right\}\oplus \mbox{span}\;\left\{{{z}^{k}\otimes {e}_{i}/k\ge
0,i=0,\ldots \ell }\right\} \\
\noalign{}
& \quad & \oplus \; \mbox{span}\;\left\{{{
z}^{k}\otimes {f}_{i}/k>0,i=0,\ldots \ell }\right\}
\end{eqnarray*}

Let $\omega = \omega_c$ and let $\Lambda :{b}^{st}\rightarrow \rlap/C $ be a
1-dimensional representation of $b^{st}$ defined by
$$
\Lambda \left({{e}_{i}}\right)=0 \quad \Lambda \left({{
h}_{i}}\right)={m}_{i}\in {Z}_{+} \quad \left({i=0,\ldots ,
\ell }\right)
$$

These representations are called the integrable highest weight representations. In
particular if $g$ is finite-dimensional, these are the finite dimensional
representations. The fundamental weights $\Lambda_0, \Lambda_1, \ldots 
,\Lambda_l$ are
such that
$$
\left.\begin{array}{lll}
{\Lambda }_{j}\left({{H}_{k}}\right) & = &{2\left({{\Lambda
}_{j},{\alpha }_{k}}\right) \over \left({{\alpha }_{k},{\alpha
}_{k}}\right)}={\delta }_{jk}\\ 
{\Lambda }_{j}\left({d}\right) &= &0 
\end{array}\right\} \quad j,k=0,\ldots ,\ell
$$

In this way given the fundamental weights of a finite dimensional Lie algebra {\it \.g}
$$
{\dot{\Lambda }}_{j}\left({{H}_{k}}\right)={2\left({{\dot{\Lambda
}}_{j},{\alpha }_{k}}\right) \over \left({{\alpha }_{k},{\alpha
}_{k}}\right)}={\delta }_{jk}\quad j,k=1,\ldots ,\ell 
$$
we can construct the fundamental weights of the Kac-Moody algebra $g$ as extensions 
of
the fundamental weights $\dot{\Lambda }$ in the following way:
$$
\begin{array}{ll} 
{\Lambda }_{j} &= {\dot{\Lambda }}_{j}+{\mu }_{j}{\Lambda
}_{0} \\ 
{\mu }_{j} &= -\sum\nolimits\limits_{k=1}^{\ell } {A}_{ok}
{\left({{({\dot{A}})}^{-1}}\right)}_{kj}
\end{array}
$$

{\noindent \spaceskip=0.25em $\dot A$ being the Cartan matrix of $\dot g$ and $\Lambda_0$ the linear functional
defined in paragraph 1.}

Every integrable highest weight representation is unitarizable.

\section{Elementary representations}

We know that for a finite dimensional simple Lie algebra $\dot g$ an infinite
dimensional highest weight representation is unitarizable only if $\dot \omega$ is a
consistent antilinear anti-involution corresponding to a hermitian symmetric space
(see [7],[8]).

The remaining unitarizable representations can be constructed only for Kac-Moody
algebras related to these type of finite dimensional Lie algebras.

 {\spaceskip=0,25em Let $ {\dot b} =$ \hskip -7truemm
\vtop{\hsize=2,1truecm {\box1 $\hskip 6truemm\oplus$} \par \vskip -3truemm \box2 ${\scriptstyle \alpha
\in \dot{\Delta}_+\cup\left\{0\right\}}$} $\dot{g}_\alpha$ be a Borel subalgebra of the
finite dimensional Lie algebra $\dot g$.}

Consider the parabolic subalgebra (called ``natural'')
$$
{p}^{\rm nat}={z}^{n}\otimes \dot{b}= {\rm span}\left\{{{z}^{n}
\otimes {h}_{i},{z}^{n}\otimes {e}_{i}}\right\}, \quad n \in
Z, \quad i=1, \ldots ,\ell
$$

Take a Cartan decomposition of the Lie algebra $\dot g$ corresponding to a hermitian
symmetric space:
$$
\dot{g}={\dot{p}}^{-}\oplus \dot{k}\oplus {\dot{p}}^{
+}, \quad \dot{k}={\dot{k}}^{-}\oplus \dot{\eta }\oplus {\dot{
k}}^{+}$$
where $\dot p$ and $\dot k$ are the subspace and the subalgebra corresponding to
non-compact and compact roots respectively.

We define an antilinear anti-involution $\omega$ of $g$ by
$$
\left.\begin{array}{ll} 
\omega \left({{z}^{n}\otimes {k}_{i}^{+}}\right) &={
z}^{-n}\otimes {k}_{i}^{-} \quad i=2,\ldots ,\ell \\ 
\omega\left({{z}^{n}\otimes {p}_{ 1}^{+}}\right) &={-z}^{-n}\otimes {p}_{1}^{-}\\ 

\omega \left({{z}^{n}\otimes {h}_{i}}\right) &={z}^{-n}\otimes {h}_{i}
\quad i=0,1,\ldots ,\ell 
\end{array}\right\} \quad n \in Z
$$
where $k_2^+, \ldots , k_l^+$ belong to ${\dot k}^+$ and where $p_1^+$ belongs to the
root space corresponding to the unique simple non compact root. In the previous
notation $e_1=p_1^+$ and $e_i=k_i^+$ for $i = 2, \ldots , \ell$.

Consider now a set of highest weights $\Lambda_1, \ldots ,\Lambda_N$ corresponding 
to unitarizable highest weight modules for the hermitian symmetric space $\dot g$ 
(for an explicit calculation see [1],[2]). Define a representation 
${p}^{\rm nat}\rightarrow \rlap/C $ by
$$
\Lambda \left({{z}^{k}\otimes x}\right)
=\sum\nolimits\limits_{i=1}^{N} {C}_{i}^{k}{\Lambda }_{i}
\left({x}\right)
$$
for $x \in \dot b$ and ${C}_{k}^{i}\in \rlap/C $ with $\left|{{
C}_{k}^{i}}\right|=1$.

Then the resulting representation is called ``elementary'' and it is unitarizable.

\section{Exceptional representations}

Another class of unitary representations (called ``exceptional'') are constructed in
this paragraph for the Kac-Moody algebra ${z}^{k}\otimes su\left({
n,1}\right) \quad k\in Z, \quad n\ge 1$.

Let $\dot{g}=su\left({n,1}\right)$ and let be a Cartan
decomposition $\dot{g}=\dot{k}\oplus \dot{p}$. Then $\dot{h
}={\dot{k}}_{1}\cap \dot{h }\oplus R{\dot{
h}}_{c}$ where ${\dot{k}}_{1}=\left[{\dot{k},\dot{k}}\right]$
and $\dot{h}_c$ belongs to the center of $\dot{g}$.

We take a realization of $g={z}^{k}\otimes su\left({n
,1}\right)$ in terms of matrices $\left({{a}_{ij}\left({z}\right)}\right)
\quad i,j=0,\ldots ,n$. The matrix elements are of the form $a
\left({z}\right)=\sum\nolimits\limits_{n\in Z} {a}_{n}
{e}^{in\theta }$ with $z={e}^{i\theta }$. We will use the notation
$\overline{a}\left({z}\right)=\sum\nolimits\limits_{n\in 
Z} {\overline{a}}_{n}{e}^{in\theta }$. Let $p=\left\{{\left({{
a}_{ij}\left({z}\right)}\right)\in g/{a}_{ij}=0 \quad {\rm if}\;
i>j}\right\}$ be the parabolic subalgebra. The antilinear anti-involution
acts in this case as
$$
\begin{array}{lll}
\omega \left({{z}^{k}\otimes {h}_{c}}\right) &:
&\omega \left({{a}_{00}\left({z}\right)}\right)={\overline{
a}}_{00}\left({{z}^{-1}}\right) \\ 
\omega \left({{z}^{k}\otimes {\dot{p}}^{+}}\right) &: &\omega \left({{a}_{
0j}\left({z}\right)}\right)={-\overline{a}}_{j
0}\left({{z}^{-1}}\right) \quad j=1,\ldots ,n \\ 
\omega\left({{z}^{k}\otimes {\dot{k}}_{1}}\right) &: &\omega 
\left({{a}_{ij}\left({z}\right)}\right)={\overline{a}}_{ji}
\left({{z}^{-1}}\right) \quad i,j=1,\ldots ,n\
\end{array}
$$

Define a representation $\Lambda :p\rightarrow \rlap/C $ by
$$
\begin{array}{lll} 
\Lambda \left({{a}_{ij}\left({z}\right)}\right) &= &0 \qquad i,j = 1,\ldots, n \\ 
\Lambda \left({{a}_{0j}\left({z}\right)}\right) &= &0 \qquad j = 1,\ldots ,n \\
\Lambda\left({{a}_{00}\left({z}\right)}\right)
&= &-\int_{{s}^{1}}{a}_{00}\left({{e}^{i\theta }}\right)d\mu \left({\theta
}\right) = -\varphi \left({{a}_{00}\left({z}\right)}\right) 
\end{array}
$$
where $\mu \left({\theta }\right)$ is a positive Radon mesure defined in the
unit circle $s^1$ and infinitely supported.

It can be shown (see [3]) that the hermitian form $H$ (we remind that it is completely
determined by giving $w, p$ and a representation of $p$) is positive definite and then
the corresponding representation in the space $L_{p,w}(L)$ is unitary.

\section{Tensor products}

The only remaining possibility in order to complete the set of all unitarizable
representations for affine Kac-Moody algebras is the corresponding to the highest
component of a tensor product of an elementary with an exceptional representation
for ${z}^{k}\otimes su\left({n,1}\right)$ (see [4]).

\bigskip
Explicit results concerning the
construction of these unitary representations will be given in a forthcoming paper.


\begin{thebibliography}{00} 

\bibitem{} Garc{\'\i}a-Escudero, J. and Lorente, M.: `Highest Weight Unitary Modules
for Non-Compact Groups and Applications to Physical Problems',  {\it Symmetries in
Science V} (ed. B. Gruber et al.) Plenum Publishing. New York (1991)

\bibitem{} Garc{\'\i}a-Escudero, J. and Lorente, M.: `Classification of Unitary Highest
Weight Representations for Non-compact Real Forms'. {\it J. Math. Phys.}, $781-790$ (1990).

\bibitem{} Jakobsen, H.P. and Kac, V.: `A new class of unitarizable highest weight
representations of infinite dimensional Lie algebras', in {\it Non-Linear Equations in
Classical and Quantum Field Theory} (ed. N. S\'anchez). Lect. Notes in Phys.  226,
Springer Verlag (1985).

\bibitem{} Jakobsen, H.P. and Kac, V. `A new class of unitarizable highest weight
representations of infinite dimensional Lie algebras II'. {\it J. Funct. Anal.} 82 (1989).

\bibitem{} Kac, V.G. Infinite dimensional Lie algebras; Cambridge University Press 
third edition, Cambridge 1990 .

\bibitem{}  Cornwell, J.F. Group Theory in Physics, Vol III. Academic Press (1989).

\bibitem{} Jakobsen, H.P. `Hermitian Symmetric Spaces and their Unitary Highest 
Weight Modules'. {\it J. Funct. Anal.} $52, 385-412$ (1983) .

\bibitem{} Enright, T; Howe, R.; Wallach, N. `A classification of Unitary Highest Weight
Modules, in Representation Theory of Reductive Groups' (ed. P. Trombi) {\it Progress in
Math.} 40. Birkha\" user, Boston (1983).

\end{thebibliography}
\end{document}